\documentclass[aps,pra,twocolumn,groupedaddress, showpacs]{revtex4-1}
\usepackage{graphicx}
\usepackage{amsmath}
\usepackage{amssymb}
\usepackage{siunitx} 
\DeclareSIUnit\gauss{G}

\usepackage{xcolor}
\usepackage[hidelinks]{hyperref}
\hypersetup{pdfpagelayout={TwoPageRight}}

\usepackage{comment}

\begin{document}

\title{Trapping of ultra cold atoms in a $^3$He/$^4$He dilution refrigerator}

\author{F.~Jessen}
\author{M.~Knufinke}
\author{S.\,C.~Bell}\email{sbell@pit.physik.uni-tuebingen.de} 
\author{P.~Vergien}
\author{H.~Hattermann}
\author{P.~Weiss}
\author{M.~Rudolph}
\author{M.~Reinschmidt}
\author{K.~Meyer}
\author{T.~Gaber}
\author{D.~Cano}
\author{A.~G\"unther}
\author{S.~Bernon} \altaffiliation[Present adress: ]{Quantronics group, SPEC (CNRS URA 2464), IRAMIS, DSM, CEA-Saclay, 91191 Gif-sur-Yvette, France}
\author{D.~Koelle}
\author{R.~Kleiner} \author{J. Fort\'agh}\email{fortagh@uni-tuebingen.de} 

\affiliation{CQ Center for Collective Quantum Phenomena and their Applications in LISA$^+$, Physikalisches Institut, Eberhard-Karls-Universit\"at T\"ubingen, Auf der Morgenstelle 14, D-72076 T\"ubingen, Germany}

\begin{abstract}
	We describe the preparation of ultra cold atomic clouds in a dilution refrigerator. The closed cycle $^3$He/$^4$He cryostat was custom made to provide optical access for laser cooling, optical manipulation and detection of atoms. We show that the cryostat meets the requirements for cold atom experiments, specifically in terms of operating a magneto-optical trap, magnetic traps and magnetic transport under ultra high vacuum conditions. The presented system is a step towards the creation of a quantum hybrid system combining ultra cold atoms and solid state quantum devices.
\end{abstract}

\pacs{37.10.Gh 07.20.-N}

\maketitle

The development of cold atom/solid state hybrid systems holds the promise of creating a quantum interface between printed electronic circuits, atoms and light~\cite{Wallquist09,Xiang13,Sorensen2004,Rabl2006,Petrosyan2008,Verdu2009,Petrosyan2009,Patton13,Henschel2010,Hafezi2012} with applications in quantum electronics and information processing. Several groups are currently preparing cold atomic clouds in the vicinity of superconducting chips at \SI[]{77}[]{\kelvin}, cooled by liquid nitrogen, and at \SI[]{4}[]{\kelvin}, cooled by liquid $^4$He~\cite{Willems95, Nirrengarten06, Mukai07, Mueller2010a, Cano08, Kasch2010, Cano2011, Emmert09, Bernon2013,Roux08,Shimizu2009,Nogues09,Zhang2012}. The vision of quantum state transfer between superconducting circuits and cold atoms requires further experimental development, in particular the preparation of atomic clouds close to millikelvin surfaces. This low temperature is required to operate superconducting quantum circuits and also enhances the coherence time of solid state quantum bits, which must be long enough to realize quantum state transfer to atomic degrees of freedom.

The conditions for cold atom preparation and the operation of mK environments are, however, very different. The first requires several tens of milliwatts of laser power for laser cooling~\cite{Metcalf99}. The second is highly sensitive to heat sources such as laser radiation, since the cooling power of dilution refrigerators is typically less than a milliwatt. Here, we describe an experimental setup that fulfills the requirements for both the production of ultra cold atoms and operation of a mK environment. We trap rubidium atoms in a \SI[]{6}[]{\kelvin} environment inside a $^3$He/$^4$He dilution refrigerator capable of mK temperatures. We demonstrate the operation of a magneto-optical trap loaded by a beam of slow atoms produced with a Zeeman slower. The MOT coils and end section of the Zeeman slower are constructed with superconducting electromagnets mounted on an additional \SI[]{6}[]{\kelvin} plate of the cryostat. We transfer the atoms into a magnetic trap and demonstrate the first step in a magnetic transfer scheme to bring the atoms towards the mK environment.


\section{Cryostat and vacuum system}
\begin{figure*}[ht]
	\centerline{\includegraphics[width=\linewidth]{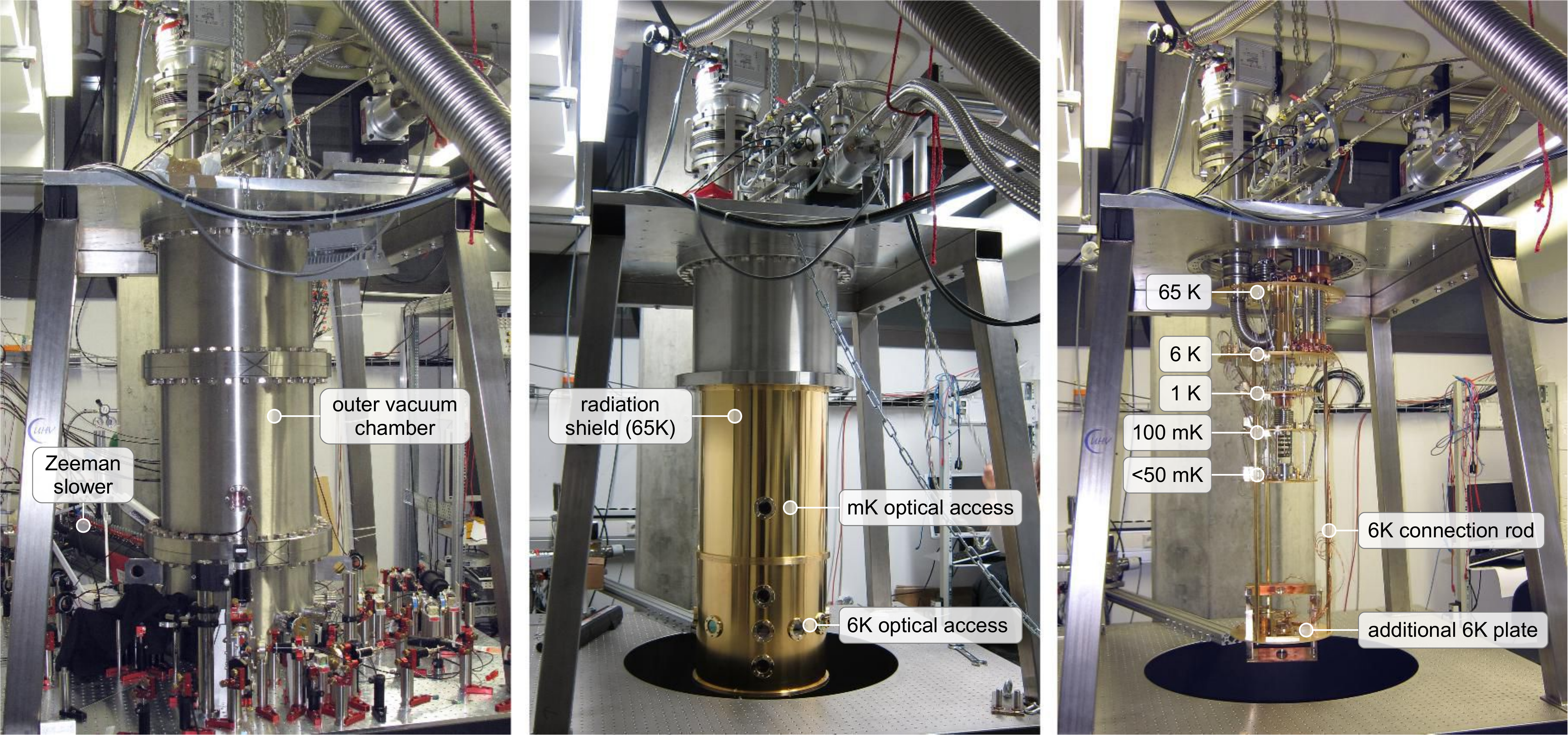}}
		\caption[Photograph of cryostat]{(Color online) Photograph of the cryostat. Left: System during operation, the cryostat is installed in the stainless steel outer vacuum chamber (OVC) and optics for the cold atom experiments surround the chamber. Center: After removal of the OVC, the \SI[]{65}[]{\kelvin} radiation shield is visible with the optical viewports: at the lower part for cold atom preparation in the \SI[]{6}[]{\kelvin} stage and the upper viewport for cold atom detection in the mK stage. Right: All radiation shields unmounted, showing the different temperature stages of the cryostat, including the additional \SI[]{6}[]{\kelvin} plate with the cold atom preparation setup mounted.}
	\label{fig:cryostat_photo}
\end{figure*}

\begin{figure}[bt]
	\centerline{\includegraphics[width=\linewidth]{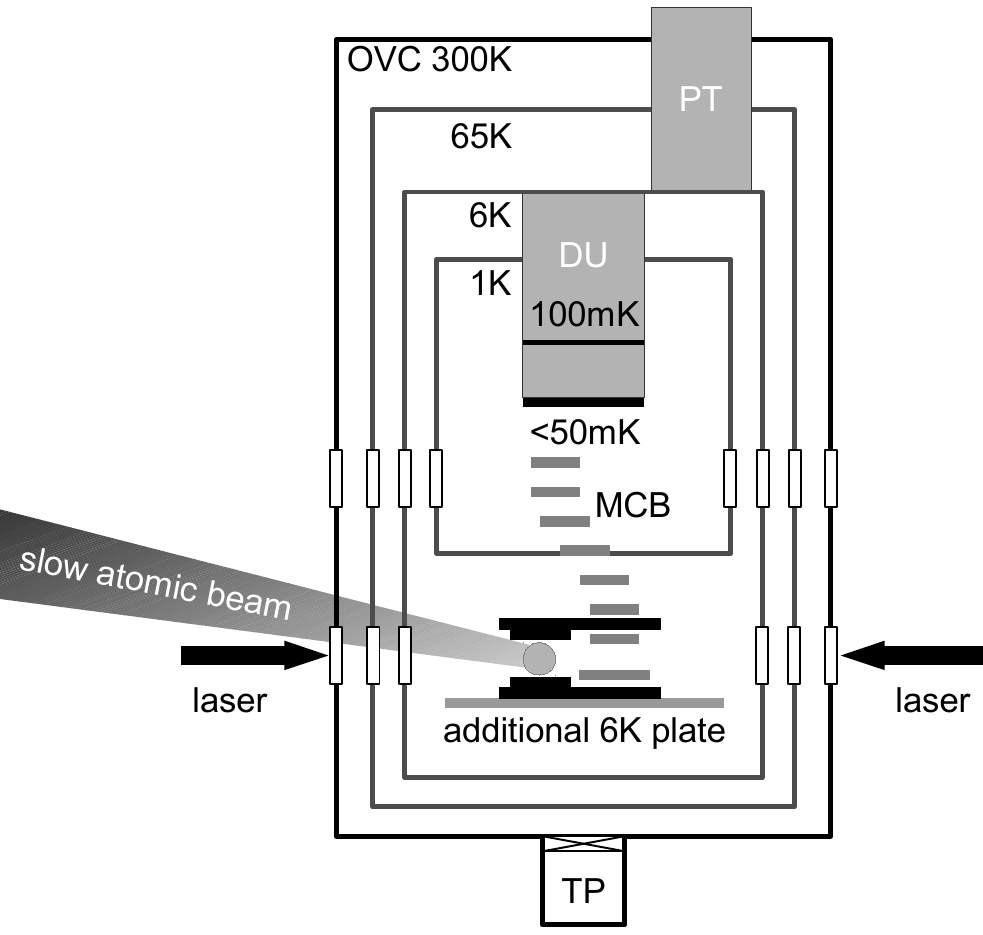}}
		\caption[Schematic of cryostat]{Schematic of the dilution refrigerator indicating the cooling stages and radiation shields. The slow atomic beam and optical access ports through the radiation shields are indicated, along with a schematic of the electromagnets for cold atom preparation mounted on the additional \SI[]{6}[]{\kelvin} plate. The conveyor belt to take atoms from the \SI[]{6}[]{\kelvin} stage to the mK is also indicated. It follows an intentionally curved path to avoid \SI[]{6}[]{\kelvin} thermal radiation reaching the mK environment. OVC: outer vacuum chamber, PT: pulse tube, DU: dilution unit, TP: turbo-molecular pump and MCB: magnetic conveyor belt.}
	\label{fig:cryostat_scheme}
\end{figure}

Cryogenic temperatures are achieved using a closed cycle $^3$He/$^4$He dilution refrigerator, based on the Oxford Instruments Triton 200 system, shown in figure \ref{fig:cryostat_photo}. The cryostat consists of two pulse tube cooled stages with cooling power of 35~watts at nominal \SI[]{45}[]{\kelvin} and 1~watt at nominal \SI[]{4}[]{\kelvin}. Due to the heat load on these stages they operate at temperatures of \SI[]{65}[]{\kelvin} and \SI[]{6}[]{\kelvin}. The \SI[]{6}[]{\kelvin} stage cools a dilution unit which has three stages operating at temperatures of \SI[]{1.5}[]{\kelvin}, \SI[]{100}[]{\milli\kelvin} and in the final stage a temperature below \SI[]{50}[]{\milli\kelvin}. Figure~\ref{fig:cryostat_scheme} is a schematic representation of the cryostat, including the position of the cold atom preparation setup mounted on the additional \SI[]{6}[]{\kelvin} plate. As can be seen in figure~\ref{fig:cryostat_photo} the additional plate for the cold atom preparation setup is mounted below the mK stage, and three large copper rods provide thermal contact to the \SI[]{6}[]{\kelvin} stage above the dilution unit. Three gold plated copper shields at temperatures of \SI[]{65}[]{\kelvin}, \SI[]{6}[]{\kelvin} and \SI[]{1.5}[]{\kelvin}, respectively, protect each progressive cooling stage from thermal radiation. 

The \SI[]{65}[]{\kelvin} and \SI[]{6}[]{\kelvin} radiation shields have a \SI[]{15}[]{\milli\meter} diameter access port for the slow atomic beam, as well as ten optical access ports to the center of the \SI[]{6}[]{\kelvin} cold atom preparation setup for the cooling and trapping light. The optical access ports are covered by fused silica windows (viewing diameter \SI[]{35}[]{\milli\meter}) which have a transmission cutoff at approximately 
\SI[]{4}[]{\micro\meter}, thus blocking the majority of \SI[]{300}[]{\kelvin} thermal radiation. The windows are anti-reflection coated at \SI[]{780}[]{\nano\meter} for high transmission of the cooling and trapping light. Two optical access ports are also placed below the mK stage, for optical diagnosis of the cloud in the mK environment.

The cryostat is mounted inside a large stainless steel vacuum chamber, which allows for the creation of ultra high vacuum conditions, as necessary for the cold atom experiments. The vacuum conditions also provide thermal insulation between the cryostat cooling stages. The vacuum chamber was initially pumped by a turbo-molecular pump, when the system is cold the cryostat surfaces serve as a high surface area cryopump and the turbo-molecular pump is isolated from the chamber with a gate valve. A pressure of \SI[]{e-9}[]{mbar} is measured with a cold cathode gauge in close proximity to the room temperature outer vacuum chamber. Due to efficient cryopumping, the pressure in the cooling and trapping region is significantly lower, as evident by long magnetic trap lifetimes.

Thermal anchoring of the superconducting electromagnets for the cold atom preparation setup was a particularly important issue. The superconducting electromagnets were wound with single filament niobium-titanium wire (\SI[]{50}[]{\micro\meter} diameter, $T_c$=\SI[]{9.2}[]{\kelvin}) that is embedded in a copper matrix (combined diameter \SI[]{80}[]{\micro\meter}) and insulated with a Kapton layer (total diameter \SI[]{100}[]{\micro\meter}). The wires were thermally anchored to the cryostat by placing them into channels machined into copper `anchor blocks' and filled with indium. Despite these thermal anchoring efforts the measured critical current of the wires were between \SI[]{0.4}[]{\ampere} and \SI[]{0.7}[]{\ampere} (depending on the coil), which is below the \SI[]{1}[]{\ampere} used in the design and significantly below the manufacturer specified \SI[]{10}[]{\ampere}. This is presumably due to localized heating at imperfections in the superconducting wire which in a wet cryostat would be efficiently cooled by thermal contact with liquid helium. The reduced magnetic field gradients, due to limited current in the coils, have not prevented the operation of the cold atom stage, however, they have degraded its performance. To improve the performance of the cold atom preparation setup in future experiments, the single filament wires will be replaced with multi filament wires.


\section{Cold atom setup}

\begin{figure}[tb]
		\centerline{\includegraphics[width=\linewidth]{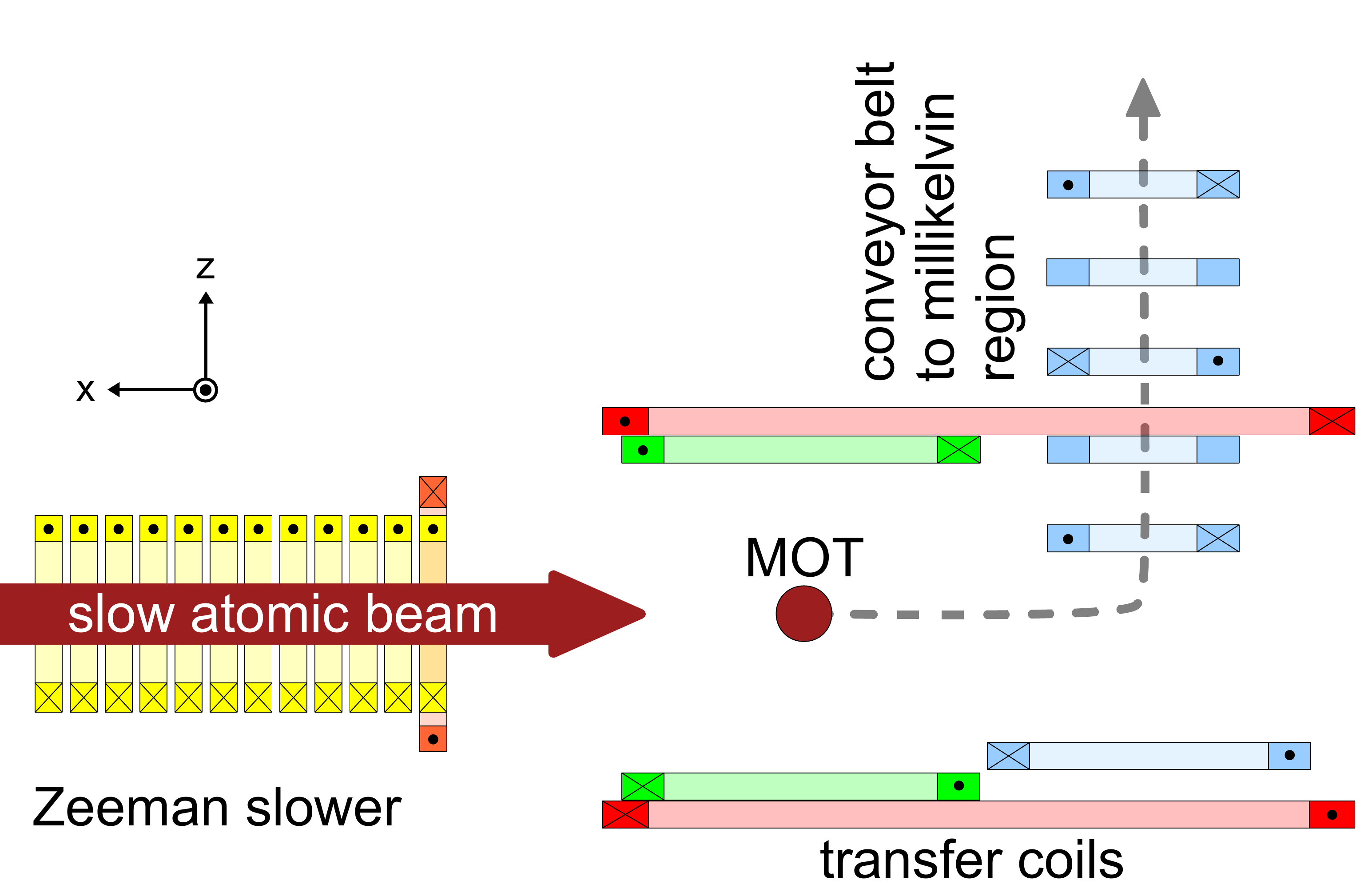}}
		\caption[Schematic of cold atom setup]{(Color online) Schematic of the cold atom preparation setup, constructed with superconducting coils on the additional \SI[]{6}[]{\kelvin} plate. The cold section of the Zeeman slower is indicated in yellow, including a compensation coil in orange. The MOT coils are in green and enclosed in the red transfer coils. The first section of the magnetic conveyor belt, which will be used to transport atoms to the mK environment is indicated in blue.}
	\label{fig:cold_atom_schematic}
\end{figure}

The preparation of cold atoms follows standard cooling and trapping techniques~\cite{Metcalf99}, with the exception of the use of superconducting electromagnets to create the required magnetic field gradients~\cite{Willems95}, which have the advantage of negligible Ohmic heating. An effusive oven, combined with a Zeeman slower~\cite{Phillips82}, creates a beam of slow atoms which are captured in a magneto-optical trap (MOT)~\cite{Raab87} and then transferred into a magnetic trap. In future experiments the cold atomic cloud will be transported from this preparation setup in the \SI[]{6}[]{\kelvin} environment to the mK environment using a magnetic conveyor belt, the path of which will be curved to avoid \SI[]{6}[]{\kelvin} thermal radiation reaching the mK environment. The first step of magnetic transfer has been realized. A schematic of the cold atom preparation setup is shown in Figure~\ref{fig:cold_atom_schematic}.

\subsection{Slow atomic beam} 

The preparation of a sample of cold atoms begins with a beam of fast atoms effusing out of a hot oven and decelerated by a Zeeman slower in a zero-crossing configuration~\cite{Bell10}. The oven is similar to the design presented in~\cite{Lin09}. The rubidium oven is typically operated at a reservoir temperature of \SI[]{80}[]{\degreeCelsius} and a collimation tube temperature of \SI[]{120}[]{\degreeCelsius}, chosen as a compromise between producing a high atomic flux and preserving the oven lifetime. A mechanical shutter at the end of the collimation tube blocks the atomic beam after loading of the MOT. 

The required magnetic field profile for Zeeman slowing is created by a series of coils, divided into two sections. The coils comprising each section run equal current and the local magnetic field is given by the number of windings in each coil. One coil section creates the positive magnetic field (i.e., from the oven until the zero-crossing). This section is at room temperature, constructed with a \SI[]{1.1}[]{\meter} long stainless steel vacuum tube (\SI[]{8}[]{\milli\meter} inside diameter) and attached to the outer vacuum chamber of the cryostat. The coils surrounding the vacuum tube are constructed with copper wire running \SI[]{5}[]{\ampere} and are water cooled to remove dissipative heat losses.

The atomic beam exits the positive Zeeman slower section and enters the cryostat through a series of \SI[]{15}[]{\milli\meter} holes in the radiation shields. The beam subsequently enters the negative section of the Zeeman slower (after the zero-crossing), which runs \SI[]{-0.3}[]{\ampere} and is constructed with superconducting wire and mounted on the additional \SI[]{6}[]{\kelvin} plate. The performance of the negative section is limited due to the reduced current capabilities of the superconducting wires, necessitating modification of the Zeeman slower operating parameters.

Rubidium-87 atoms with an initial velocity below \SI[]{335}[]{\meter\per\second} are decelerated along the Zeeman slower to a final velocity of \SI[]{24}[]{\meter\per\second} over a distance of \SI[]{1.23}[]{\meter} by a slowing laser beam counter propagating to the atomic beam. The magnetic field at the beginning of the slower is \SI[]{250}[]{\gauss} and the exit magnetic field is \SI[]{-35}[]{\gauss}, where positive is defined as the propagation direction of the slowing laser. Slowing is achieved using \SI[]{25}[]{\milli\watt} of $\sigma^+$ polarized light (relative to the direction of propagation), tuned \SI[]{80}[]{\mega\hertz} below the \mbox{5S$_\text{1/2} F$=2} $\rightarrow$ \mbox{5P$_\text{3/2} F^\prime$=3} cycling transition. An additional \SI[]{8}[]{\milli\watt} of re-pump light, tuned \SI[]{80}[]{\mega\hertz} below the \mbox{5S$_\text{1/2} F$=1} $\rightarrow$ \mbox{5P$_\text{3/2} F^\prime$=2} transition returns any atoms that are lost from the cycling transition.

\subsection{Magneto-optical trap}
\label{sec:MOT_loading}

The exit of the Zeeman slower is located at a distance of \SI[]{55}[]{\milli\meter} from the center of the MOT, which was intentionally kept short to minimize the diffusion of atoms as they coast out of the Zeeman slower and into the trapping region. A compensation coil is placed between the end of the slower and the MOT to ensure that the magnetic field of the Zeeman slower is canceled at the center of the MOT.

The magneto-optical trap consists of three orthogonal pairs of counter-propagating cooling laser beams with a total power of \SI[]{65}[]{\milli\watt}. The Gaussian profile of the beams is cut with an aperture at \SI[]{20}[]{\milli\meter} diameter to produce a nearly homogeneous beam profile. The beams intersect at the center of a pair of superconducting coils with an inner diameter of \SI[]{22}[]{\milli\meter} and separation of \SI[]{24}[]{\milli\meter} (indicated in green in fig.~\ref{fig:cold_atom_schematic}). The coils produce a quadrapole magnetic field with a gradient of approximately \SI[]{250}[]{\gauss\per\centi\meter} in the $z$ direction, running a current of \SI[]{150}[]{\milli\ampere}. The cooling light is detuned \SI[]{17}[]{\mega\hertz} below the \mbox{5S$_\text{1/2} F$=2} $\rightarrow$ \mbox{5P$_\text{3/2} F^\prime$=3} cycling transition in $^{87}$Rb. An additional \SI[]{8}[]{\milli\watt} re-pump laser beam, on resonance with the $^{87}$Rb \mbox{5S$_\text{1/2} F$=1} $\rightarrow$ \mbox{5P$_\text{3/2} F^\prime$=2} transition is overlapped with the cooling beams and re-pumps atoms lost into the $F$=1 state back into the cooling cycle.

\begin{figure}
	\centerline{\includegraphics[height=\linewidth, angle=-90]{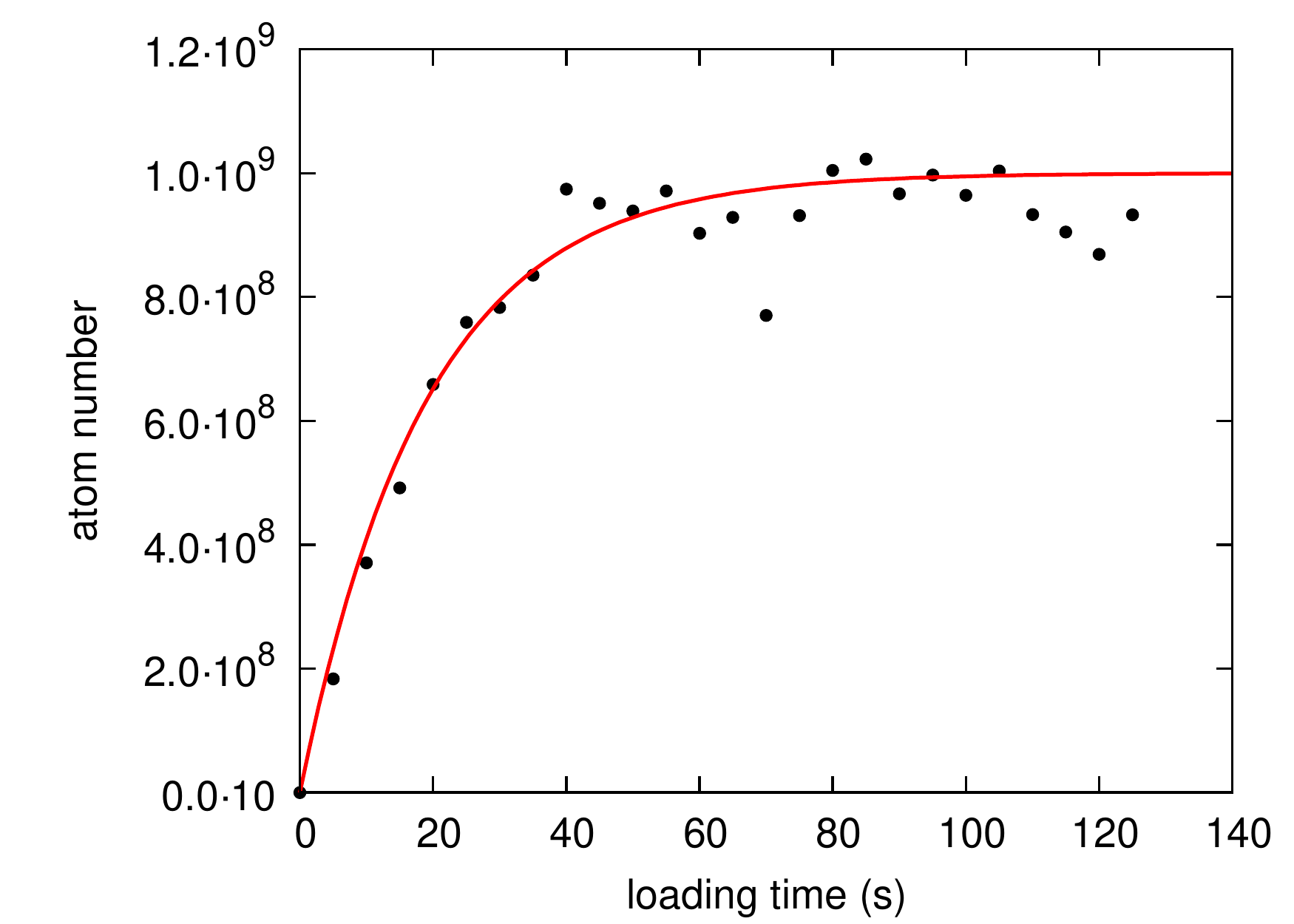}}
		\caption[Plot of MOT loading]{(Color online) A typical curve of the MOT loading from the Zeeman slower. Data points were measured by absorption imaging and the data was fit to an exponential function $N(t)=N_0 \cdot (1-\exp(-t/\tau_\text{l}))$. The fit indicates a loading time of $\tau_\text{l}=19$\,s.}
	\label{fig:MOT_loading}
\end{figure}

Figure~\ref{fig:MOT_loading} shows the number of atoms in the magneto-optical trap as a function of loading time, measured by absorption imaging~\cite{Ketterle99}. The red curve in is an exponential loading fit to the data: $N(t)=N_0 \cdot (1-\exp(-t/\tau_\text{l}))$, where $N_0$ is the steady state atom number, $t$ is time and $\tau_\text{l}$ is the loading constant. The fit indicates a loading constant of $\tau_\text{l}=19$\,s. The MOT loads \num[]{5e8} atoms in 10~seconds and saturates at \num[]{1e9} atoms after 100~seconds. Such a loading rate is considered low for a Zeeman slower system, but is not surprising given that the Zeeman slower is not running at the design specifications due to current limitation in the superconducting wire.

In a typical experimental cycle we load the MOT for 10~seconds, trapping \num[]{5e8} atoms at a temperature of \SI[]{230}[]{\micro\kelvin}. The cooling light is then detuned to \SI[]{67}[]{\mega\hertz} and the magnetic field gradients are simultaneously ramped down over \SI[]{12}[]{\milli\second}. After a further \SI[]{5}[]{\milli\second} of molasses cooling a \SI[]{250}[]{\micro\second} optical pumping pulse is applied in a weak homogeneous field to transfer atoms into the magnetically low field seeking state $F$=2, $m_F$=2. 

The magnetic field ramps used during molasses cooling and magnetic trapping are intentionally longer than \SI[]{10}[]{\milli\second}, exceeding the decay time scale of eddy currents induced in the surrounding copper coil supports and cryostat components. Future improvements to the system will reduce the eddy currents, while maintaining essential high quality thermal contact to the cryostat. 

\subsection{Magnetic trap} 

\begin{figure}
	\centerline{\includegraphics[height=\linewidth, angle=-90]{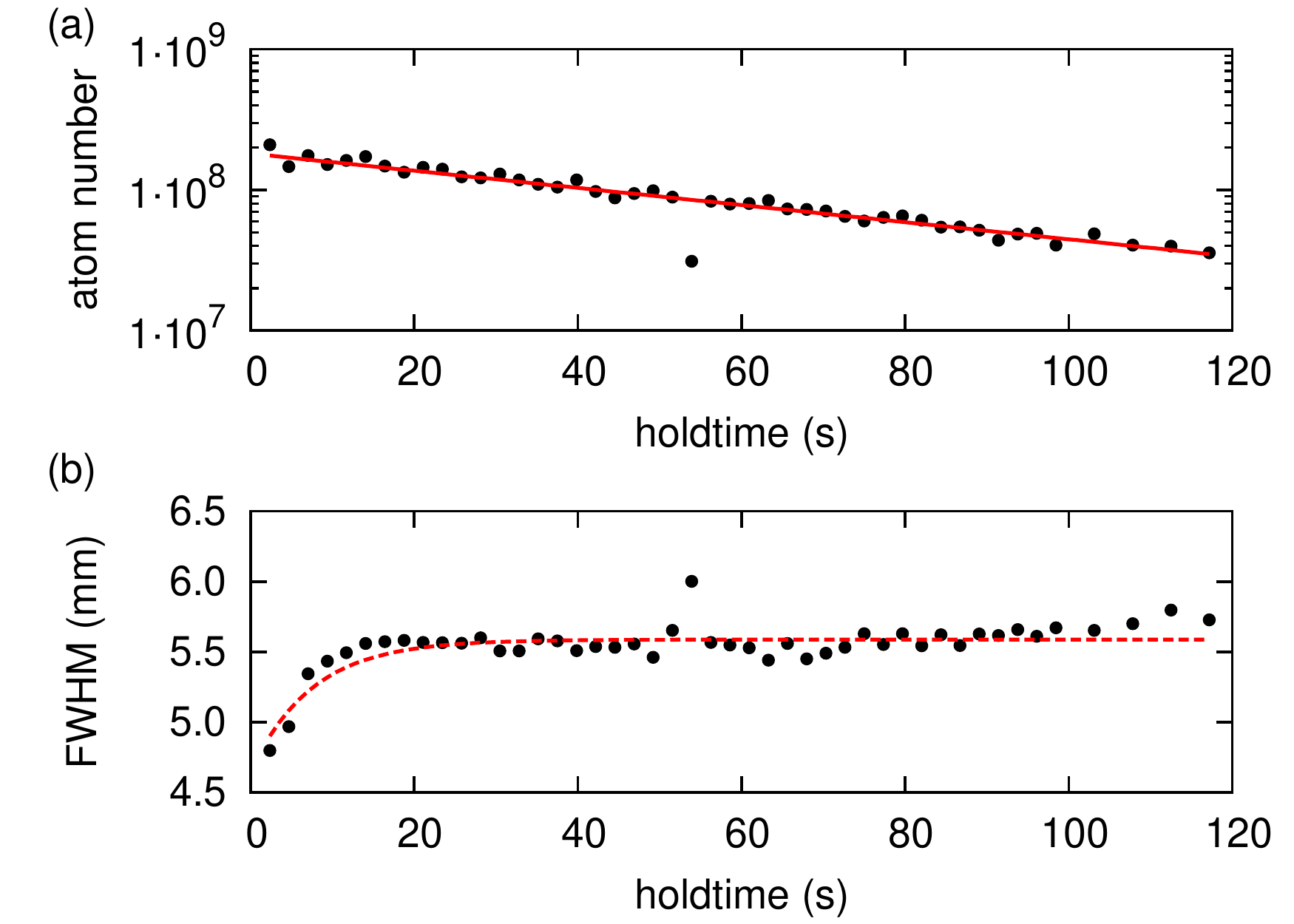}}
		\caption[Decay of atoms in the magnetic trap]{(Color online) (a) A curve of the decay of atoms out of the magnetic trap, as measured by absorption imaging. The data was fit to an exponential decay function $N(t)=N_0 \cdot \exp(-t/\tau_\text{d})$, and indicates a magnetic trap lifetime of $\tau_\text{d}=70$\,s. (b) The FWHM of the cloud in the horizontal direction over the lifetime of the cloud. The initial increase in the cloud width is consistent with a thermalization time of \SI[]{7.5}[]{\second}, as calculated for our trapping parameters and densities (red dashed curve).}
	\label{fig:B_decay}
\end{figure}

We transfer \num[]{e8} atoms into a magnetic quadrupole trap by increasing the current in the MOT coils to \SI[]{500}[]{\milli\ampere}, resulting in a field gradient of approximately \SI[]{30}[]{\gauss\per\centi\meter} in the $z$ direction. The temperature of the ensemble in the magnetic trap was measured to be \SI[]{90}[]{\micro\kelvin} by ballistic expansion in time-of-flight images~\cite{Vorozcovs05}. Figure~\ref{fig:B_decay} (a) is a logarithmic plot of the atom number as a function of hold time in the magnetic trap, indicating the trap lifetime. The data is well described by an exponential decay: $N(t)=N_0 \cdot \exp(-t/\tau_\text{d})$ with a time constant of $\tau_\text{d} =70$\,s. Plotted in figure~\ref{fig:B_decay}(b) is the full width half maximum (FWHM) of the cloud size in the horizontal direction over the lifetime of the trap. The initial increase in the FWHM of the cloud is consistent with a thermalization time scale of \SI[]{7.5}[]{\second}, as calculated for our trapping parameters and densities (red curve). 

We have observed a long term decrease in the trap lifetime which is related to an increase in the vacuum chamber pressure. The pressure increase was first detected after many months of operation of the system, supporting the hypothesis that the pressure increase is due to the saturation of the cryopumping surfaces~\cite{Mundinger1992}. To provide optimal vacuum conditions we regularly condition the cryopumping of the system by allowing the cold atom preparation setup to warm up to \SI[]{20}[]{\kelvin} while the system is pumped by the turbo-molecular pump~\cite{Mundinger1992}. Re-cooling the system to \SI[]{6}[]{\kelvin} improves the vacuum pressure and the magnetic trap lifetime is recovered. Similar observations have been reported in~\cite{Willems95}.

\subsection{Magnetic transfer} 

\begin{figure}
		\centerline{\includegraphics[height=\linewidth, angle=-90]{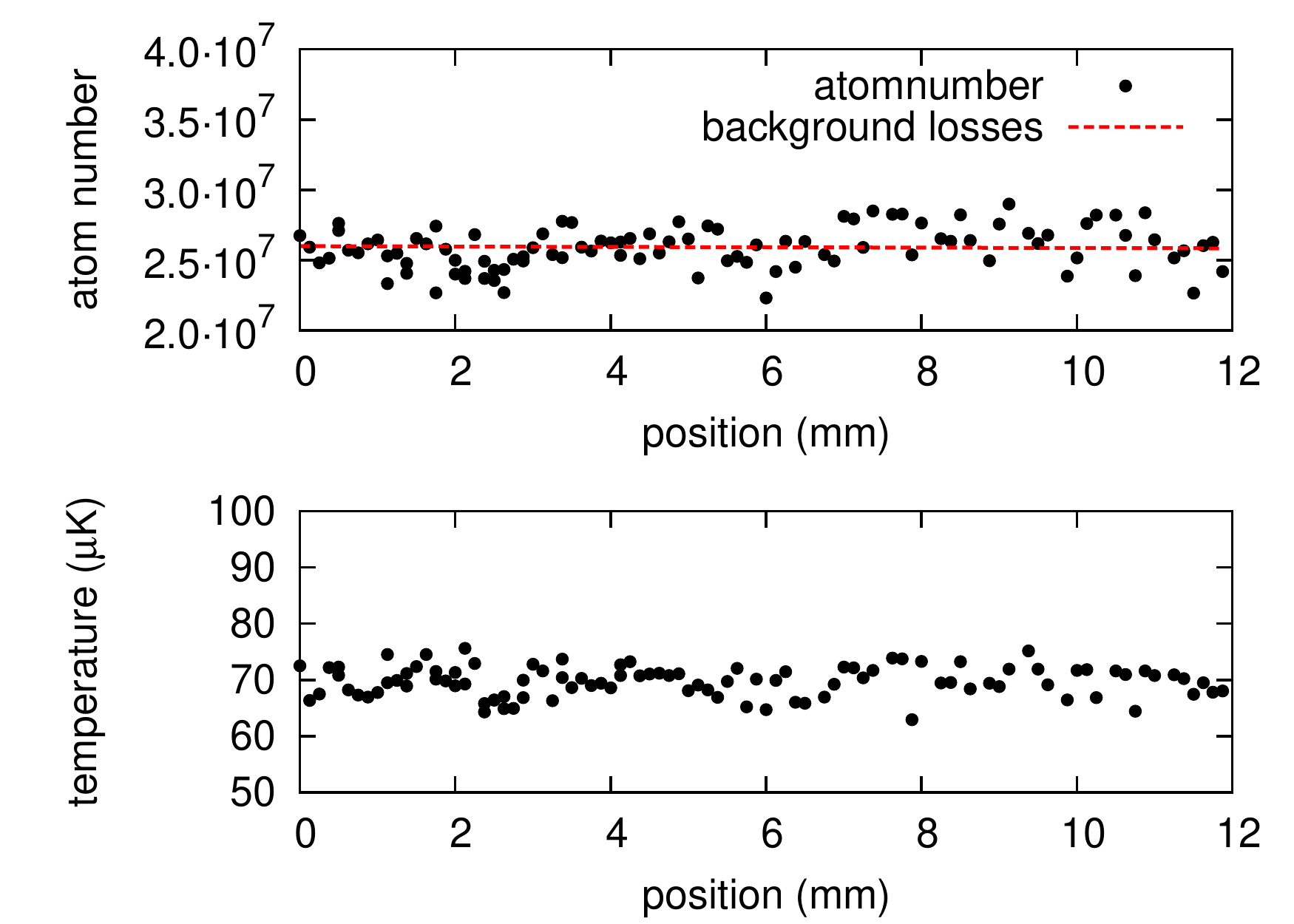}}
		\caption{(Color online) (a) Atom count during magnetic transport of the cloud, as measured by absorption imaging. The cloud was transported a varied distance from the center of the MOT coils and then back again for imaging. Accounting for background losses (red dashed curve), the data indicates that a \SI[]{70}[]{\micro\kelvin} cloud can be transferred in 1~second with nearly 100\% efficiency. (b) The temperature of the cloud during transport, showing no evidence of heating.}
	\label{fig:B_transfer}
\end{figure}

The atomic cloud is transferred from the magnetic trap created by the MOT coils into the first transfer coil pair (indicated in red in fig.~\ref{fig:cold_atom_schematic}). This transports the atomic cloud approximately \SI[]{12}[]{\milli\meter} in the horizontal direction. Transfer is achieved by increasing the current in the transfer coils, and simultaneously decreasing the current in the MOT coils. A transfer time of 1~second was found to optimize the number of transferred atoms. To assess the transfer efficiency the cloud was transported a varied distance from the center of the MOT coils and then back again for imaging diagnostics (figure~\ref{fig:B_transfer}). The slight atom loss evident in the data can be explained by the magnetic trap lifetime, as indicated in the dashed red curve. 

Taking these background losses into account, we find a transfer efficiency of nearly 100\% for an initial cloud temperature of \SI[]{70}[]{\micro\kelvin}. For higher temperatures, the efficiency of the transfer is limited by the trap depth when the magnetic quadrupoles of the two coil pairs merge. 

This magnetic transport will play a major role in the next step in our experiments, in which atoms will be transported from the \SI[]{6}[]{\kelvin} cold atom preparation setup to the mK environment. The full magnetic conveyor belt structure has been designed and is now being installed.


\section{Conclusion}

We have prepared an ensemble of cold atoms in a closed cycle $^3$He/$^4$He cryostat. The magnetic field gradients required for laser cooling and trapping are produced entirely with superconducting electromagnets, which carry a current of approximately \SI[]{0.5}[]{\ampere} in the \SI[]{6}[]{\kelvin} stage of a dilution refrigerator. We have realized the first step in a magnetic transfer scheme in which future experiments will transport a cold atomic cloud to a mK environment, where we will bring the atoms into contact with superconducting solid state devices. 

This apparatus provides the required experimental conditions for the realization of hybrid quantum systems \cite{Verdu2009,Petrosyan2009} combining ultracold gases and superconducting quantum devices, and also for fundamental studies on ultra cold Rydberg atoms in absence of detrimental thermal radiation \cite{Amthor2009}.

\section{Acknowledgments}
This work was supported by the European Research Council (ERC Advanced Grant SOCATHES) and the Deutsche Forschungsgemeinschaft (SFB TRR21).
M.K. and M.R. acknowledge support from the Carl Zeiss Stiftung.
P.V. und H.H. acknowledge support from the Evangelisches Studienwerk e.V. Villigst.

\bibliography{bibliography}

\end{document}